# Effects of metallic absorption and the corrugated layer on the optical extraction efficiency of organic light-emitting diodes

Baek-Woon Lee[1], Young-Gu Ju[2*]
*[1]Samsung Mobile Display, San 24, Nongseo-dong, Giheung-gu, Yongin-si, Gyeonggi-do, 449-711, Korea*
*[2]Department of Physics Education, KyungPook National University, 702-701, Daegu, Korea*

The absorption of a metallic cathode in OLEDs is analyzed by using FDTD calculation. As the light propagates parallel to the layer, the intensity of $E_z$ polarization decreases rapidly. The intensity at 2.0 μm from the dipole is less than a quarter of that at 0.5 μm. The strong absorption by a cathode can be a critical factor when considering the increase of optical extraction by means of bending the optical layers. The calculation indicates that the corrugation of layers helps the guided light escape the guiding layer, but also increases the absorption into a metallic cathode. The final optical output power of the corrugated OLED can be smaller than that of the flat OLED. On the contrary, the corrugated structure with a non-absorptive cathode increases the optical extraction by nearly two times.

## 1. INTRODUCTION

Organic light-emitting diodes (OLEDs) have been studied due to their many advantages such as a low power consumption, high contrast ratio and high speed operation. Those favorable features of OLEDs may open the way to various applications, for example, flat panel displays and general lighting sources.

However, low optical efficiency is one of the obstacles to overcome in order to survive the competition with liquid crystal displays and semiconductor-based light-emitting diodes. The poor optical extraction efficiency is mainly attributed to the total internal reflection coming from the higher index of emission layers than those of an anode and a glass substrate. For a typical OLED structure, the extraction efficiency is only ~20 %[1,2].

Various efforts have been made to improve the optical extraction[3-5]. Most approaches deal with patterns inside or on the surface of the device in order to reduce internal reflections. Photonic crystal structures were also proposed and demonstrated successfully, even though the fabrication technique is not mature enough to scale up to a large area device. Such ideas basically assume that a large portion of the generated light is guided along the emission layer, and that the modification of a flat profile through some roughness can lead to giving out the light into escape modes. However, this viewpoint ignores the fact that there exists a highly absorptive metallic layer called a cathode in an OLED structure.

In this paper, the authors analyze the portion of energy absorbed into a metallic cathode by using the finite-difference-time-domain (FDTD) method. In addition, the authors checked the possibility of increasing the extraction efficiency by depositing the layers on the patterned substrate.

## 2. SIMULATION DETAILS

As mentioned earlier, the authors chose the FDTD in order to study the absorption effect of a metallic cathode as well as the corrugated layers. The FDTD is being widely used these days to solve the Maxwell equation in regards to the field of wireless communication and nano-photonics[6,7]. In principle, the FDTD is accurate, even at a scale limit smaller than a wavelength, as far as numerical errors and quantum effects are ignorable. Even though the OLED layers analyzed in this paper can be as small as a few nano-meters, they can be treated as a thicker layer together with neighboring layers, without affecting the final calculation results, because the light cannot distinguish a superlattice from an alloy. Therefore, the FDTD approach is presumed to give good numerical data, which cannot be obtained by applying the ray tracing theory. In regard to the simulation, we used a three-dimensional (3D) FDTD with periodic boundary conditions in the y direction, therefore saving a significant amount of computation time and enabling the scanning of a wide parameter range, compared to nominal 3D calculations. The periodic 3D FDTD assumes that a physical situation is symmetrical along one axis in terms of optical structure and dipole sources. For instance, if one simulates a square profile in this quasi-2D model, it corresponds to an infinitely long square pillar in the y direction. Therefore, the periodic FDTD modeling is an efficient tool for analyzing the effect of the cathode and the scattering layer in OLEDs with accuracy and speed.

An example of the permittivity profile is shown in Fig. 1. The OLED, in the example, is composed of 5 layers. From the top, they are a 200 nm Aluminum cathode, 40 nm electron transport layer plus electron injection layer, a 50 nm emission layer, a 60 nm hole transport layer plus hole-injection layer, and a 90 nm anode. The permittivity of the background is set to 2.50, which is close to that of glass. In order to analyze the absorption effect of a cathode, a dipole is placed in the emission layer and the detector circles are positioned at various distances from the dipole, as illustrated in Fig. 1. Along the detector line, the Poynting vector is measured to calculate the angular distribution of radiation and the total output power. As the radius of the detector circle increases, the total output power may vary, depending on the absorption of a cathode.

As for the polarization, the dipoles are excited in three different directions at the same time with random phase differences between themselves. The $E_x$ polarization and $E_z$ polarization refer to the horizontal and the vertical direction in the plane of Fig. 1, respectively. The $E_y$ polarization is perpendicular and directed into the is normal to the plane that is made by x and y direction.

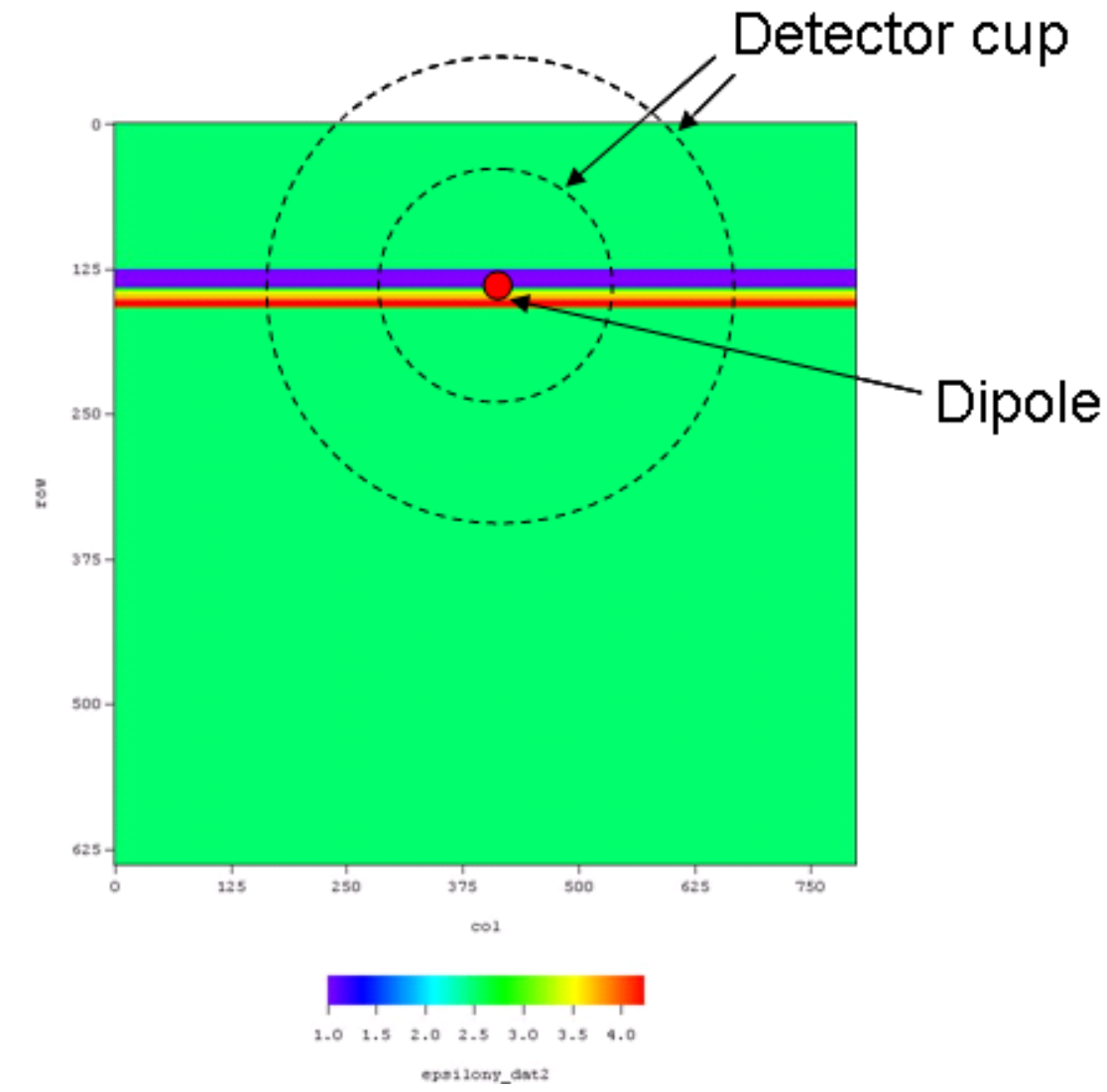

Fig. 1. The cross-section of the permittivity profile used in the calculation is displayed. A dipole is placed in the emission layer and the detector line is positioned at various distances from the source.

## 3. RESULTS AND DISCUSSION

The calculated angular distributions of dipole radiations, as seen in Fig. 2, exhibit distinctive behavior as the distance of the detector line from the source increases. The intensity peak located near +/- 90 degrees represents the guided energy between the OLED layers. Among the three polarizations, the $E_z$ polarization decays rapidly as the light propagates along the cathode. When the detector radius changes from 0.5 μm to 1.0 μm, the intensity peak at the edge drops to less than a half. A similar amount of shrinkage is also observed as the radius increases from 1.0 μm to 2.0 μm. In comparison, $E_x$ polarization and $E_y$ polarization are both insensitive to the variation of the detector radius. This may be ascribed to the fact that $E_x$ and $E_y$ polarizations mainly relate to the propagation normal to the surface while the $E_z$ polarization comprises the propagation mode parallel to the surface. $E_z$ polarization also has the intensity minimum near 0 degree or the surface normal since a dipole doesn't radiate into the direction of its own oscillation.

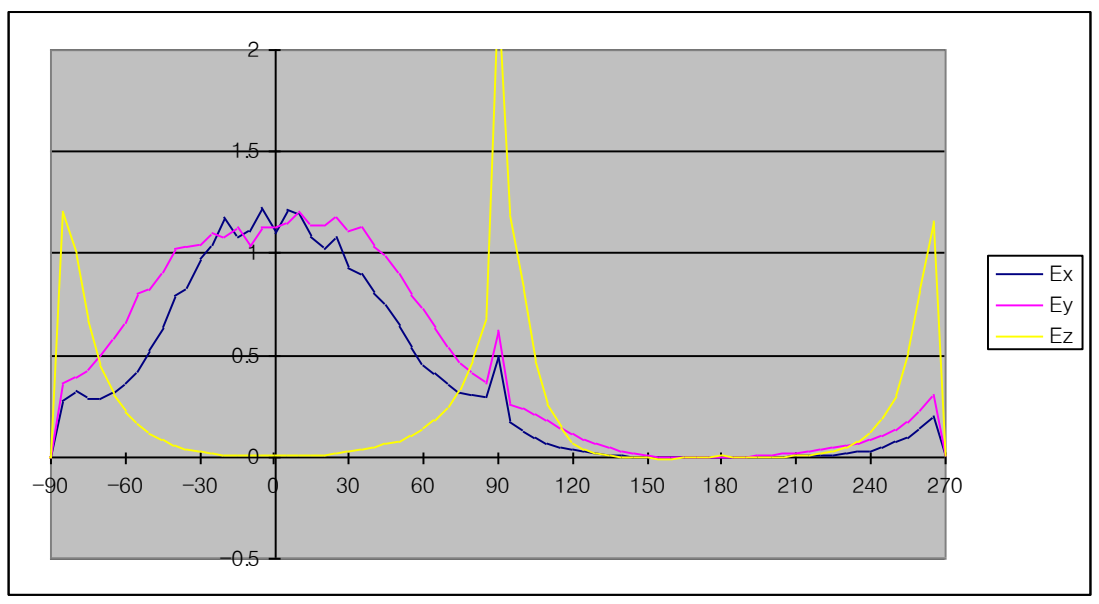

(a)

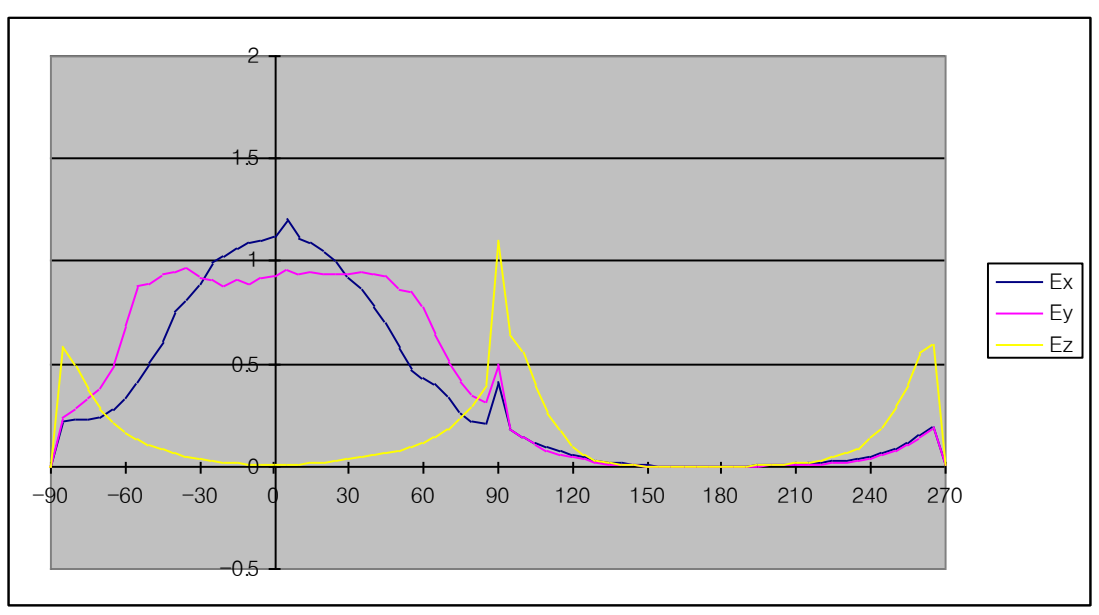

(b)

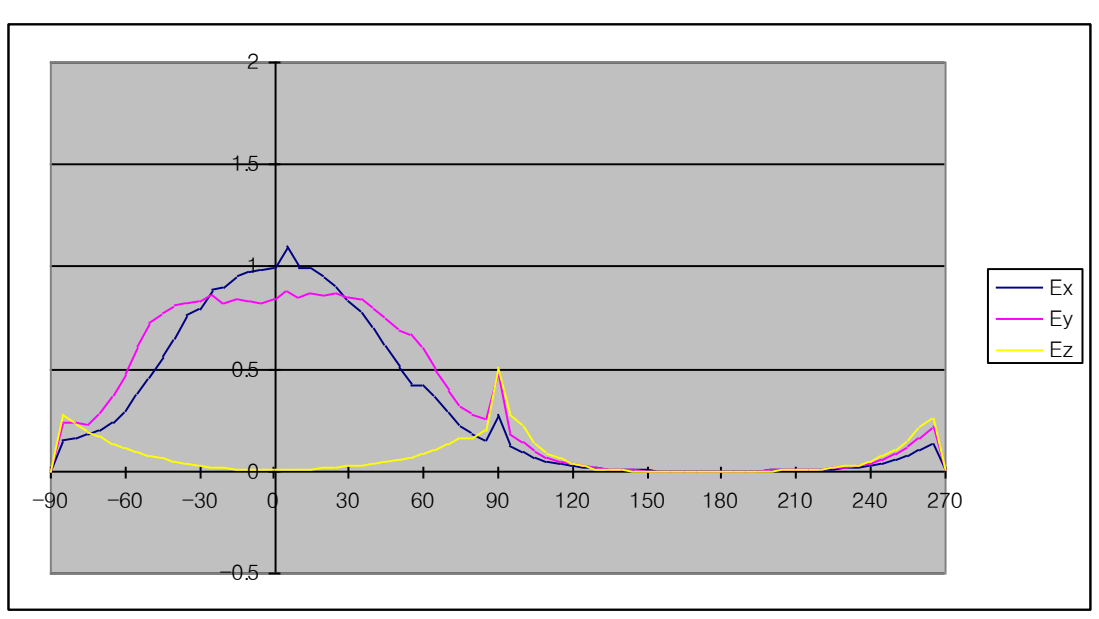

(c)

Fig. 2. Angular distribution of radiation, as measured at different distances from the dipoles in the OLED structure; (a) 0.5 μm from the source, (b) 1.0 μm from the source and (c) 2.0 μm from the source

Even though metal seems to have high reflectivity, this is due to the index difference between the metal and its neighboring material. Aluminum, used as a cathode, has a refractive index of 0.644 + I 5.28 at the wavelength of 460 nm. The large imaginary index can make aluminum work as an absorbent, as well as a reflector. From this viewpoint, it is not a surprising that the cathode absorbs the light energy of the guided mode in the OLED. However, it can be a critical issue if someone intends to increase extraction efficiency by giving some patterns to the layers and inducing a disturbance in the guided mode. If the pattern is larger than a decay length as determined by a cathode, the guided energy cannot escape, because it has been already absorbed into a cathode. For example, the bending of wave guide may turn the guided mode into an escape mode. However, the bending radius should be smaller than 0.5 μm, otherwise the light energy turns into heat in a cathode before escaping.

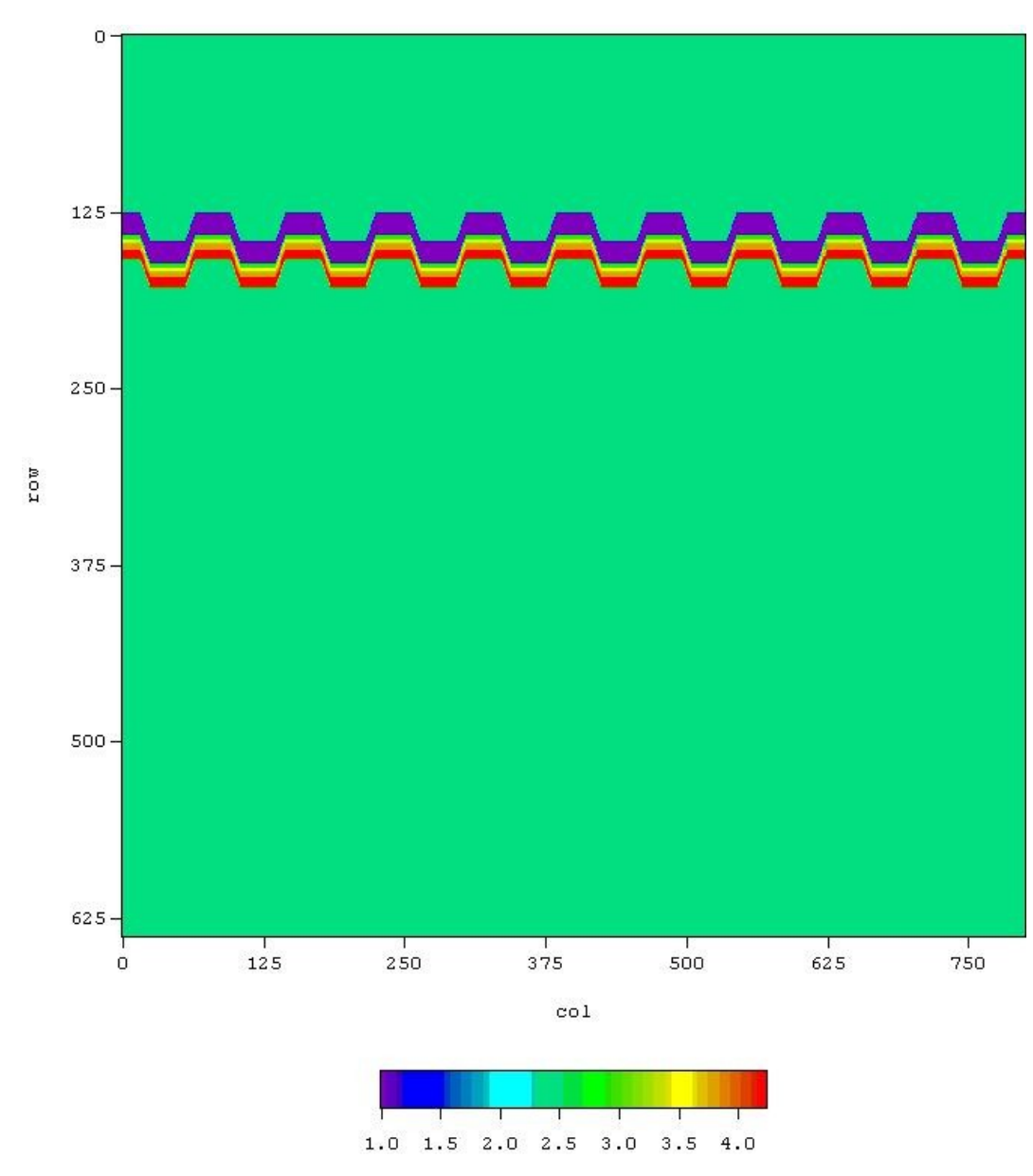

Fig. 3. The permittivity profile of the corrugated OLED is displayed. The period, the width and the height of grooves are 0.50 μm, 0.25 μm and 0.25 μm, respectively.

Along this line of thought, the authors tested a corrugated structure to perturb the guide mode, hoping to increase some extraction efficiency, as seen in Fig. 3. These corrugated layers tend to bend the wave-guiding layers to extract the guided energy, as if the bending of fiber induces the light to smear out, due to the breaking of total internal reflections.

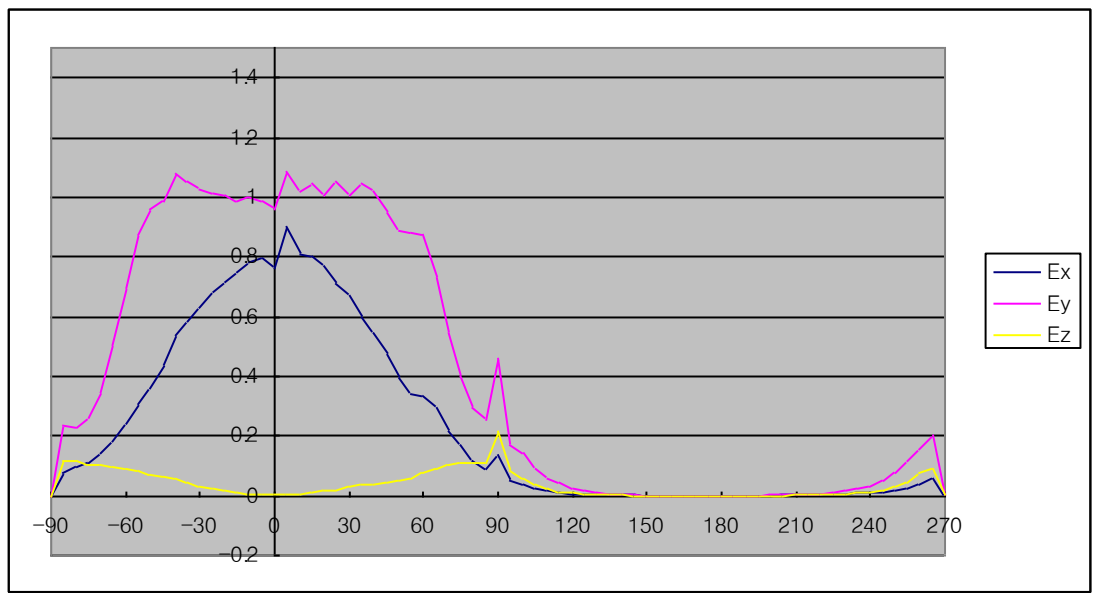

(a)

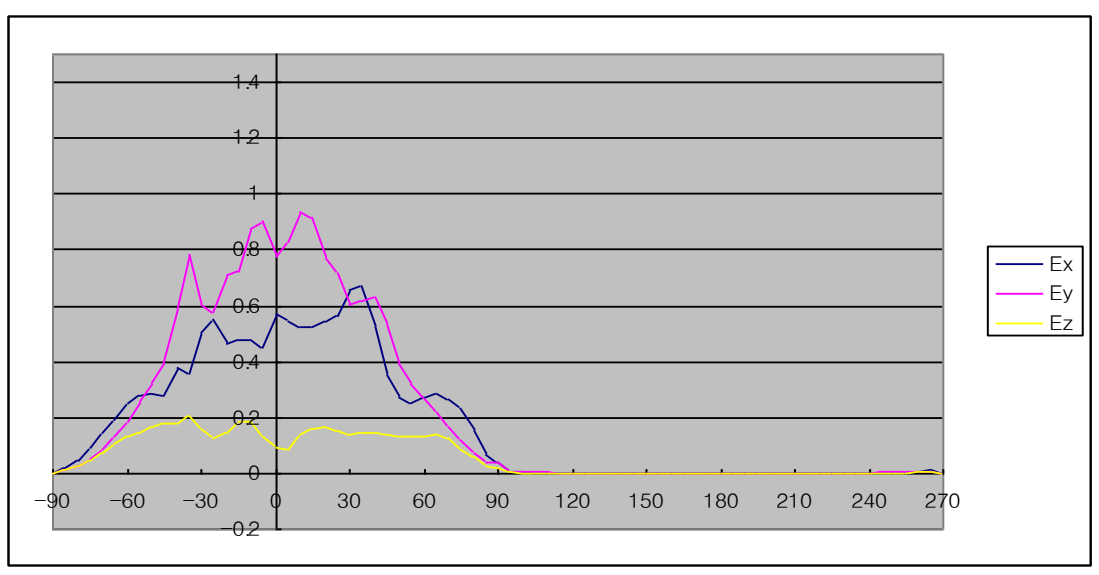

(b)

Fig. 4. Angular radiation distribution of an OLED with a metallic cathode; (a) a flat structure and (b) a corrugated structure, in which the period, width and height are 0.25 μm, 0.125 μm and 0.1 μm, respectively .

The results from a flat layer and a corrugated structure are presented in Fig. 4. In regard to the corrugated OLED, a slight increase of the intensity emitting at a normal direction is found for $E_z$ polarization, in comparison to that of the flat structure. If radiation, within 40 degrees, can escape a glass substrate, the ratio of the escaping power to the total power increases from 38 % to 51 % when the corrugation profile is used in place of the flat layer. On the contrary, the total optical output power of the corrugated OLED is only 60 % of its flat counterpart. It implies that the corrugation may act, not only to extract the energy from the guided structure, but also to enhance its metallic absorption by extending the light path along the cathode. Several other parameter ranges have been investigated to look for the feasibility of increasing the total extraction. However, the results are not so much different from those as shown in Fig. 4.

In order to confirm the effects of metallic absorption as well as the corrugated layers, the high index dielectric material is assumed to comprise the cathode. The permittivity of 900 is used for the cathode instead of the index of aluminum. This artificial permittivity may help to find the physical mechanism of the corrugated structure and the role of the metallic cathode since dielectric material reflect the light without absorption, due to the absence of an imaginary index.

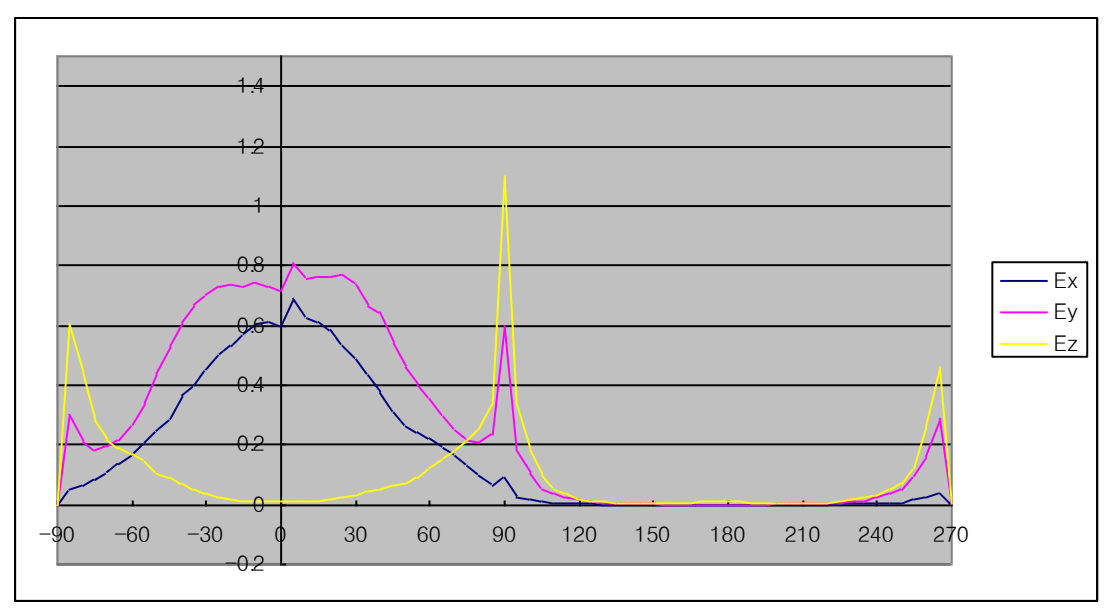

(a)

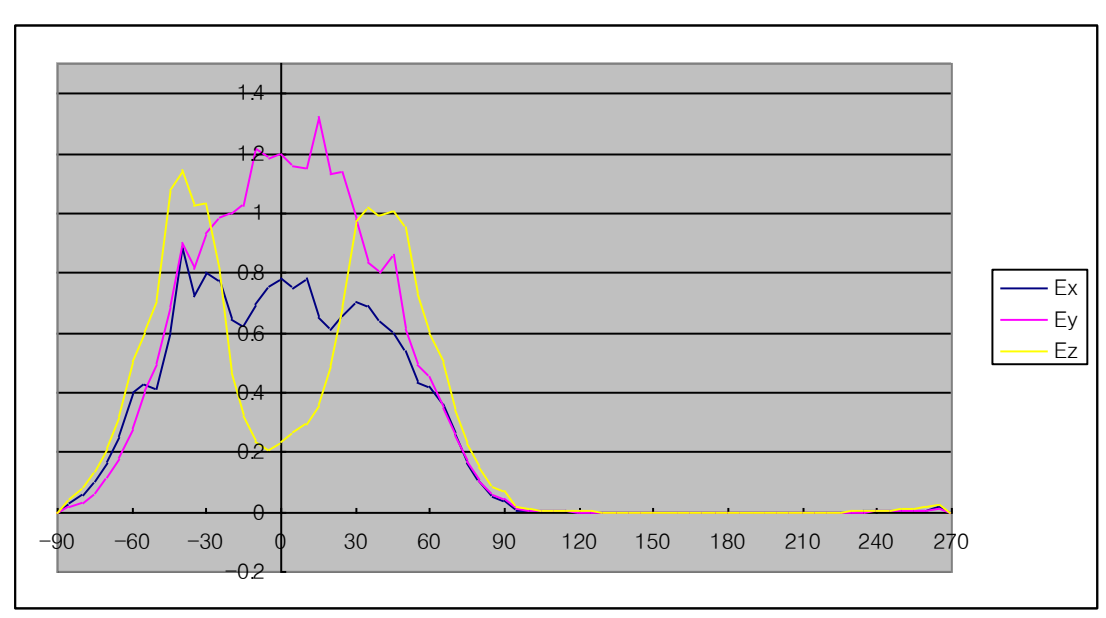

(b)

Fig. 5. Angular radiation distribution of an OLED with a non-absorptive cathode; (a) a flat structure and (b) a corrugated structure, in which the period, width and height are 0.25 μm, 1.25 μm and 0.1

μm, respectively.

In the case of the corrugated OLED with a non-absorptive cathode, the optical output within the escape cone almost doubles, in comparison to that of the flat one. It means that the corrugation can be a useful method to increase the optical extraction of OLEDs, only if the cathode doesn't absorb the light. This metallic absorption may explain the limited success of the structure, which has an emission layer coated over the photonic crystal patterned substrate[3]. Therefore, the corrugation of the emission layers requires careful consideration with respect to the metallic absorption and the risk of short-circuit, as well as a scalable fabrication technique for more practical applications.

## 4. CONCLUSION

The absorption of a metallic cathode in OLEDs is analyzed by using FDTD calculation. As the light propagates parallel to the layer, the intensity of $E_z$ polarization decreases rapidly. The intensity at 2.0 μm, apart from the dipole, is less than a quarter of that at 0.5 μm. The strong absorption by a cathode can be a critical factor to consider when trying to increase the optical extraction by means of bending the emission layers. The simulation result indicates that the corrugation of layers does not only enable the guided light to escape the guiding layer, but also increases its absorption into a metallic cathode. The final optical output power of the corrugated OLED turns out to be smaller than that of the flat OLED. The calculations regarding the structure with a non-absorptive cathode support the effects of a cathode. The purely real permittivity of a non-absorptive cathode increases the optical extraction by nearly 100 %, therefore, vindicating the importance of the metallic absorption of a cathode when designing a new optical extraction scheme.

**Acknowledgment:** This work was supported by the Korea Research Foundation Grant funded by the Korean Government (MOEHRD, Basic Research Promotion Fund) (KRF-2007-331-D00318). This work was supported by Samsung Electronics Corporation. This work was supported by the Regional Innovation Center Program (ADMRC) of the Ministry of Knowledge Economy, Republic of Korea.